\documentstyle[aps,preprint,eqsecnum]{revtex}

\begin{document}

\newcommand{\nn}{\nonumber}

\setcounter{page}{1}
\draft

\bigskip
\bigskip

\title{
ON  UNCONSTRAINED $SU(2)$ GLUODYNAMICS WITH THETA ANGLE}

\author{A.M. Khvedelidze\ $^{a,b}$, \,\,  D.M. Mladenov \ $^a$, \,\,
H.-P. Pavel\ $^{a,c}$,  and \,\, G. R\"opke $^c$}
\address{$^a$
Bogoliubov Laboratory of Theoretical Physics,
Joint Institute for Nuclear Research, 141980 Dubna, Russia}
\address{$^b$
Department of Theoretical Physics,
A. Razmadze Mathematical Institute, GE-380093 Tbilisi, Georgia}
\address{$^c$ Fachbereich Physik der Universit\"at Rostock,
D-18051 Rostock, Germany}
\date{January 23, 2002}

\maketitle

\begin{abstract}
The Hamiltonian reduction of classical $SU(2)$ Yang-Mills field theory
to the equivalent unconstrained theory of gauge invariant local dynamical
variables is generalized to the case of nonvanishing $\theta$-angle.
It is shown that for any $\theta$-angle the elimination of the pure gauge
degrees of freedom leads to a corresponding unconstrained nonlocal theory
of self-interacting second rank symmetric tensor fields,
and that the obtained classical unconstrained gluodynamics
with different $\theta$-angles are canonically equivalent as on the
original constrained level.
\end{abstract}

\section{Introduction}

The gauge- and Poincare invariant action of Yang-Mills theory
depends on two parameters, the coupling constant
$g$ and so-called $\theta$-angle,
as coefficients in front of
the CP even part $S^{(+)}$
\begin{equation}
\label{eq:action}
S^{(+)}  =  \frac{1}{2g^2}\ \int d^4x \,
\mbox{tr}\, F_{\mu\nu} F^{\mu \nu}~,
\end{equation}
and the CP odd part $S^{(-)}$
\begin{equation} \label{eq:CPodd}
S^{(-)}\, =\,
-\frac{\theta}{16\pi^2} \, \int d^4x \, \mbox{tr}
\, ^{\ast}F_{\mu\nu}  F^{\mu\nu}\,,
\end{equation}
respectively.
At the classical level neither the value of the coupling constant
nor that of the $\theta$-angle effect the observables, because the
complete information for the description of the classical
behaviour of the gauge fields is coded entirely in the extremum of the action.
When all components of the gauge potential entering the action are varied as
independent variables the topological charge density term
$Q(x)\, =\,-(1/16\pi^2)\, \mbox{tr}
\, ^{\ast}F_{\mu\nu} F^{\mu\nu} \, $
can be discarded as a total divergence
\begin{equation}\label{eq:topcharge}
\, Q(x) \, = \,
\partial_\mu K^\mu\,,
\end{equation}
with the Chern-Simons current $K^\mu$ \cite{Deser}
\begin{equation}\label{K}
K^\mu \, = \, -\frac{1}{16\pi^2}\, \varepsilon^{\mu\alpha\beta\gamma}\mbox{tr}
\left(
F_{\alpha\beta} A_\gamma - \frac{2}{3} A_\alpha A_\beta A_\gamma
\right)\,
\end{equation}
and thus the extremal curves are independent of both the coupling constant
and the $\theta$-angle.

Passing to the quantum theory it is generally believed
\cite{JackiwRebbi,Callan,Jackiw}
that the physical observables become $\theta$-dependent.
Although in perturbative calculations all diagrams with vertex \(Q(x)\)
vanish, nonperturbative phenomena such as
tunneling between the above topologically distinct
classical vacua, labeled by the integer value of the winding number functional
\begin{equation}\label{clctr1}
W[A] \, = \, \int d^3 x\, K^0\,,
\end{equation}
leads to the appearance of $\theta$-vacua.
Configurations with different winding number are related to each other by
large gauge transformations reflecting the fact that
the topological current $K_\mu$ is not gauge invariant.

We therefore pose at this place the question whether it is possible
to express the topological term in the classical action
as a total divergence of a gauge invariant current using the unconstrained
formulation of gauge theories \cite{GoldJack}-\cite{Majumdar:2001qh}.
In the hope to obtain such a representation of the topological term
we would like to generalize in the present notes the
Hamiltonian reduction of classical $SU(2)$ Yang-Mills
field theory given in \cite{KP} to arbitrary $\theta$-angle
by including the CP odd part (\ref{eq:CPodd}) of the action.
We shall reformulate the original degenerate Yang-Mills theory
as an unconstrained nonlocal theory of selfinteracting
second rank symmetric tensor fields.

Carrying out such a reduction in the presence of a total divergence term
in the action one can meet so called
 {\em ``divergence problem''} specific
for the field theory with constraints which has no analog
for finite-dimensional mechanical systems.
This problem has first been formulated explicitly in the
context of the canonical reduction of General Relativity.
\footnote{Presumably,
the idea of the importance of the careful consideration of
terms which are total spatial divergences goes back to
P.Dirac in 1959 when he constructed  the
reduced Hamiltonian in general relativity as a certain surface integral
at spatial infinity \cite{Dirac}.}
Forty years ago R. Arnowitt, S. Deser and C.W. Misner \cite{ADM} gave
a clear and vivid formulation of the phenomenon:
{\em `` a term which in the original Lagrangian ( or Hamiltonian )
is a pure divergence, may cease to be a divergence upon elimination of the
redundant variables and hence may contribute to the equations of motion
obtained from the reduced Lagrangian ( Hamiltonian )''}.
A simple {\it ad hoc} example from \cite{ADM} explains the
idea of this statement.
Consider a theory where among the variables
there is a redundant variable satisfying the constraint
\begin{equation}\label{eq:exmpl}
\nabla^2 \Phi = \chi^2\,.
\end{equation}
 A term
$
\nabla^2 \Phi
$
added to the degenerate Lagrangian being a divergence has no influence
on the classical equation of motion, while after projection onto
the constraint shell it appears as $\chi^2$ and would contribute to
equations of motion.

We shall demonstrate that the
Hamiltonian reduction of $SU(2)$ Yang-Mills gauge
theory is free of the above mentioned divergence problem
due to the Bianchi identities.
Equivalence of constrained and unconstrained formulations of gauge
theories on the classical level requires the demonstration of
the agreement between reduced and original non-Abelian Lagrangian equations
of motion. We shall explicitly construct the canonical
transformation, well defined on the reduced phase space,
that  eliminates the $\theta$-dependence of the classical equations of motion
for the unconstrained variables.\footnote{A similar
construction for gravity has been done recently in \cite{Montesinos}.}

\section{Theta independence on the constrained level}
\label{sec:cedg}

Let us first review the case of the original constrained theory
and demonstrate that under the special boundary conditions
for the fields at spatial infinity (see Eq.
(\ref{eq:boundary}) below) there exists a canonical transformation
which completely eliminates the $\theta$-dependence from
the classical degenerate theory.

\subsection{Hamiltonian fomulation of the constrained theory}

Both parts of action  $S=S^{(+)}+S^{(-)}$  are invariant under
the local gauge transformations
\begin{equation}  \label{eq:tran}
A_\mu \,\,\, \rightarrow \,\,\,  A_\mu^{\prime}   =
U^{-1}(x) \left( A_\mu  - \partial_\mu \right) U(x)~,
\end{equation}
with an arbitrary  space-time depended element $U(x)$ of
the gauge group.
This means that the Lagrangian theory is degenerate and the standard
Hamiltonian description needs to be generalized.
We shall follow the Dirac Generalized Hamiltonian
approach \cite{DiracL,HenTeit}.

Inclusion of the CP odd part of the action $S^{(-)}$ leads to the
modification of the canonical momenta
\begin{eqnarray}  \Pi_a    = \frac{ \partial{ L} }{ \partial
{\dot A}_{a0} }&=& 0  \\
  \Pi_{ai} = \frac{ \partial{ L} }{
\partial {\dot A}_{ai} } &=& \frac{1}{g^2}\left( {\dot A}_{ai}
-\left(D_{i}(A)\right)_{ac}A_{c0} \right) +
\frac{\theta}{8 \pi^2} B_{ai}~~
\end{eqnarray}
where the covariant derivative $D_i$
reads
\begin{equation}\label{eq:newcon}
\left(D_i(A)\right)_{mn} = \delta_{mn}\ \partial_i +
\left( J^c \right)_{mn}\ A_{ci}\,,
\end{equation}
with the $ 3 \times 3 $
matrix generators of $ SO(3) $ group,\\
$\left( J_s\right)_{mn} := \epsilon_{msn}$,
and non-Abelian magnetic fields
\begin{equation}
B_{ai} = \varepsilon_{ijk}\left(\partial_j A_{ak}+
{1\over 2}\epsilon_{abc}A_{bj} A_{ck}\right)\,
\end{equation}
has been introduced. Independently of this modification the phase space
spanned by
the variables $(A_{a0}, \Pi_a)$ and  $(A_{ai}, \Pi_{ai})$
is restricted by the three primary constraints
$\Pi_a (x) = 0~$.

The canonical Hamiltonian is
\begin{equation} \label{eq:canham}
H_C = \int d^3x
\Bigg[\frac{g^2}{2}\left( \Pi_{ai} -
\frac{\theta}{8\pi^2} B_{ai}\right)^2
 +\frac{1}{2g^2} B_{ai}^2
+ \Pi_{ai} \left(D_i A_{0}\right)_a
\Bigg]~,
\end{equation}
where we have used that the topological charge density
$Q(x)$ can be rewritten in terms of the non-Abelian electric
and magnetic fields as
\begin{equation}
Q =  \frac{1}{8\pi^2}  F^a_{0i} B_{ai}\,.
\end{equation}

The standard way in the Hamiltonian approach to proceed further,
is to perform a partial integration in the last term in expression
(\ref{eq:canham}) for the canonical Hamiltonian
\begin{equation}
\int_{V_R} d^3x  \Pi_{ai}\left( D_i A_{0}\right)_a
= - \int_{V_R}  d^3x A_{a0} \left(D_i \Pi_{i}\right)_a
 \ \ \ +\oint_{\Sigma_R} d^2\sigma_i  A_{a0} \Pi_{ai}\,,
\end{equation}
where according to the Gauss theorem the surface integral is over
the two-dimensional closed surface covering the three-dimensional
volume $V_R$ ( for simplicity we assume that it is a ball with radius $R$ ).
Supposing that
\begin{equation} \label{eq:boundary}
\lim_{R \to \infty} \oint_{\Sigma_R} d^2\sigma_i  A_{a0} \Pi_{ai} \,=\,0\,,
\end{equation}
we obtain the non-Abelian Gauss law constraint
\begin{equation}  \label{eq:secconstr}
\left(D_i \right)_{ac}\Pi_{ic} = 0\,,
\end{equation}
as the condition to maintain the
primary constraints\\  $\Pi_a =0$ during the evolution.
According to the Dirac prescription the generator of time translation
is the total Hamiltonian
\begin{equation}
\label{eq:totham}
H_T = \int d^3x \Bigg[\frac{g^2}{2}\left(
\Pi_{ai} - \frac{\theta}{8\pi^2} B_{ai}\right)^2 +\frac{1}{2g^2} B_{ai}^2
 -A_{a0}  D_i\Pi_{ai}  + \lambda_a \Pi_a \Bigg]~,
\end{equation}
depending on three arbitrary functions $\lambda_a (x)$ and
the Poisson brackets have a canonical structure
\begin{eqnarray}
\label{eq:pbo}
&&\{A_{ai}(\vec{x},t),\Pi_{bj}(\vec{y}, t) \}=
\delta_{ab}\delta_{ij}\delta^3(\vec{x}-\vec{y})\,,\\
&&\{A_{a0}(\vec{x},t),\Pi_{b}(\vec{y}, t) \}=
\delta_{ab}\delta^3(\vec{x}-\vec{y})\,.
\end{eqnarray}

\subsection{Canonical equivalence of constrained theories with different
 theta-angles}

Based on the representation (\ref{eq:totham}) for the total Hamiltonian
one can immediately verify
the equivalence of classical theories with different value of parameter
$\theta$.
To convince let us perform the transformation
to  new coordinates  $A_{ai}$ and $E_{bj}$
\begin{eqnarray} \label{eq:cantrtheta}
A_{ai}(x) &\to & A_{ai}(x) = A_{ai}(x)\,,\\
\Pi_{bj}(x) &\to& E_{bj} = \Pi_{bj}(x) - \frac{\theta}{8\pi^2} B_{bj}(x)\,.
\end{eqnarray}
One can easily check that this transformation is canonical, the new
coordinates $A_{ai}$ and ${E_{ai}}$ satisfy the same canonical Poisson
brackets relations (\ref{eq:pbo}) as the original one.
And noticing that by virtue of the Bianchi identity
\begin{equation}
\epsilon^{\mu\nu\lambda\rho}D_\nu F_{\lambda\rho}\,=\,0\,,
\end{equation}
one can conclude that the $\theta$-dependence completely
disappears from the Hamiltonian (\ref{eq:totham}).

Note that the canonical transformation (\ref{eq:cantrtheta})
can be represented in the form
\begin{eqnarray} \label{clctr}
E_{ai} & = & \Pi_{ai} - \theta \frac{\delta}{\delta A_{ai}}
W[A]\,,
\end{eqnarray}
where  $W[A]$ denotes the winding number functional (\ref{clctr1}).

\section{Theta independence on the unconstrained level}

We shall now derive the unconstrained version of Yang-Mills theory
with $\theta$-angle and then give the analog of the transformation
(\ref{eq:cantrtheta}) after projection to the reduced phase space,
thus checking the consistency of the unconstrained canonical formulation
of Yang-Mills theory.

\subsection{Hamiltonian formulation of the unconstrained theory}

For the reduction of $SU(2)$ Yang Mills theory we shall follow
the method developed in \cite{KP} for the CP even part of action.
To reduce the CP odd part one can proceed similarly.

Let us therefore perform the following point transformation to the new
set of Lagrangian coordinates
$q_j\ \ (j=1,2,3)$  and the six elements
$S_{ik}= S_{ki}\ \ (i,k=1,2,3)$ of the positive definite
symmetric $3\times 3$ matrix $S$
\begin{equation}\label{eq:gpottr}
A_{ai} \left(q, S \right) = O_{ak}\left(q\right) S_{ki} -
{1\over 2}\epsilon_{abc} \left( O\left(q\right) \partial_i
O^T\left(q\right)\right)_{bc}\,,
\end{equation}
where $O(q)$ is an orthogonal $3\times 3$ matrix parameterized by the
three fields $q_i$.

The first term in (\ref{eq:gpottr}) corresponds
to the so-called polar decomposition for arbitrary quadratic matrices.
The inclusion of the additional second term is motivated by the
inhomogeneity of the gauge transformation (\ref{eq:tran}).
\footnote{One can treat equation (\ref{eq:gpottr})
as gauge transformation to new field configuration $S(x)$ which satisfy
the so-called symmetric gauge condition $\epsilon_{abc}S_{bc}=0$.
The uniqueness and regularity of the transformation
(\ref{eq:gpottr}) depends on the boundary conditions imposed.}
The transformation (\ref{eq:gpottr}) induces a point canonical
transformation linear in the new conjugated momenta \( P_{ik} \)
and \(p_i \). Using the corresponding  generating functional
depending on the old momenta and the new coordinates,
\begin{equation}
F_3 \left[ \Pi; q, S \right] =
\int d^3z \ \Pi_{ai}(z) A_{ai} \left(q(z), S(z)\right)~,
\end{equation}
one can obtain the transformation to new canonical momenta
$ p_i $ and $ P_{ik} $
\begin{eqnarray}  \label{eq:mom1}
p_j (x)& =  & \frac{\delta F_3 }{\delta q_j(x)} =
- \Omega_{jr}\left(D_i(Q)S^T\Pi\right)_{ri}~,\\
\label{eq:mom2}
P_{ik}(x)& =  &\frac{\delta F_3}{\delta S_{ik}(x)}=
\frac{1}{2}\left(\Pi^T O + O^T \Pi \right)_{ik}~.
\end{eqnarray}
Here
\begin{equation}
\label{Omega}
\Omega_{ji}(q) \, : = \,-\frac{1}{2}
\mbox{Tr}\ \left(
O^T\left(q\right)\frac{\partial O \left(q\right)}
{\partial q_j}
\, J_i \right).
\end{equation}

The symplectic structure of new variables is encoded in the
fundamental Poisson brackets\footnote{These new brackets
take into account the symmetry constraints $S_{ij}=S_{ji}$
and $P_{kl}=P_{lk}$ and rigorously speaking are
the Dirac brackets.}
\begin{equation} \label{eq:Diracb}
\{S_{ij}(x),P_{kl}(y)\}= \frac{1}{2}\left(\delta_{ik}\delta_{jl} +
\delta_{il}\delta_{jk}\right)\delta^{(3)}(x-y)\,.
\end{equation}

A straightforward calculation based on the linear relations
(\ref{eq:mom1}) and (\ref{eq:mom2})
between the old and the new momenta leads to the following expression for
old momenta $\Pi_{ai}$ in terms of the new canonical variables
\begin{equation}  \label{eq:elpotn}
\Pi_{ai} = O_{ak}\left(q\right) \biggl [\,  P_{\ ki} +
\epsilon _{kis}  P_s\,\biggr]\,,
\end{equation}
where the  vector \( P_s \) is a solution to the
system of first order partial differential equations
\begin{equation} \label{eq:forP}
{}^\ast D_{ks}(S)P_s \,=\, {s}_k (x) +
\Omega^{-1}_{kl} p_{l} \,.
\end{equation}
In (\ref{eq:forP}) the ${}^\ast D$ denotes  the  matrix operator
\begin{equation}
\label{DeltaQ}
{}^\ast D_{ik}(S)  = -\left(D_m(S)J^m\right)_{ik}~,
\end{equation}
and one can verify that vector
\begin{equation}
{s }_k (x)  = (D_i(S))_{kl}P_{il}
\end{equation}
coincides up to a divergence term with the spin density part of the
Noetherian angular momentum calculated in terms of
the new variables and projected onto the
constraint shell.
Using the representations (\ref{eq:gpottr}) and (\ref{eq:elpotn})
one can easily convince oneself that the new variables $S$ and $P$ make
no contribution to the Gauss law constraints (\ref{eq:secconstr})
\begin{equation}
O_{as}(q) \Omega^{-1}_{\ sj}(q)p_j = 0\,.
\end{equation}
Here and in (\ref{eq:elpotn}) we assume that the matrix $\Omega$ is
invertible and thus the equivalent set of Abelian constraints is
\begin{equation}
p_a = 0\,.
\end{equation}
The  Abelian form of Gauss law constraints is the main advantage of new
variables. In terms of this coordinates the projection to the
constraints shell is achieve by  vanishing value of momenta $p_a$
in all expressions.

The reduced Hamiltonian is defined as projection
of total Hamiltonian to the constraint shell
$p_a=0$ and $\Pi_a=0$. In terms of the unconstrained canonical variables
$S$ and $P$ it reads
\begin{equation} \label{eq:uncYME}
H =
\int d^3{x}
\biggl[
{g^2\over 2}\left(P_{ai} - \frac{\theta}{8\pi^2} B_{(ai)}\right)^2+
 g^2 \left(P_{a} - \frac{\theta}{8\pi^2} B_{a}\right)^2
+{1\over 2g^2} B^2_{ai}\biggr].
\end{equation}
Here $B_{(ai)}$ and $B_a$ denote the symmetric
tensor $B_{(ai)}= (B_{ai}+B_{ia})/2$ and vector
$B_a=\epsilon_{abc}B_{bc}/2$
constructed from chromomagnetic field
\begin{equation}
B_{sk} = \epsilon_{klm}
\left( \partial_l S_{sm} + \frac{1}{2}\,
\epsilon_{sbc} \, S_{bl}S_{cm} \right)~.
\end{equation}
The  vector $P_{a}$ representing the nonlocal term in the Hamiltonian
(\ref{eq:uncYME})
is given as the solution to the system of differential equations
\begin{equation}
\label{vecE}
{}^\ast D_{ks}(S) P_s = {s}_k (x)\,,
\end{equation}
which is the projection of Eqs.(\ref{eq:forP})
to the constraint surface $p_a=0$.

\subsection{
Canonical equivalence of the unconstrained theory with different
 theta angles}

For the original degenerate action in terms of the $A_{\mu}$ fields
the equivalence of classical theories
with arbitrary value of $\theta$-angle has been reviewed in Section
\ref{sec:cedg}.
Let us now examine the same problem for the derived unconstrained
theory considering the analog of the canonical transformation
(\ref{eq:cantrtheta}) after projection onto the constraint
surface
\begin{eqnarray} \label{eq:uncantrtheta}
S_{ai}(x) &\to & S_{ai}(x) = S_{ai}(x)\,,\\
P_{bj}(x) &\to& E_{bj}(x) = P_{bj}(x) - \frac{\theta}{8\pi^2} B_{(bj)}(x)\,.
\end{eqnarray}
First of all
one can easily check that this transformation
to new variables $S_{ai}$ and ${E_{bj}}$
is canonical
with respect to the Dirac brackets (\ref{eq:Diracb}).
The Hamiltonian (\ref{eq:uncYME}) in terms of the new variables
$S_{ai}$ and ${E_{bj}}$  is therefore $\theta$-independent.
It looks as
\begin{equation} \label{eq:ht1}
H =
\int d^3{x}
\biggl[\,
{g^2\over 2}E_{ai}^2
+\, g^2 E_{a}^2
+{1\over 2g^2} B^2_{ai} \ \biggr]\,.
\end{equation}
where $E_a$ is a solution to equation (\ref{vecE})
with the replacement $P_{ai} \to E_{ai}$.
This follows from the observation, that if $P_a$ is a solution
to equation (\ref{vecE}) then
expression
\begin{equation}
E_a = P_{a} - \frac{\theta}{8\pi^2} B_{a}
\end{equation}
is a solution to the same equation with
the replacement $P_{ai} \to E_{ai}$ . This
is indeed valid because $B_{ai}$ field satisfies the identity
\begin{equation}  \label{eq:bm}
{}^\ast D_{ks}(S) B_s = (D_i(S))_{kl}B_{(li)}\,.
\end{equation}
Equation (\ref{eq:bm}) is
the Bianchi identity  $(D_i)_{ab}B_{bi} =0$ rewritten in terms of
the symmetric  $B_{(ai)}$ and
antisymmetric  $B_a$ parts of the chromomagnetic field strength.

The reduced form of the generating functional (\ref{clctr1}) corresponding
to the transformation (\ref{eq:uncantrtheta}) is the same functional $W$
evaluated for the symmetric tensor $S_{ik}$.
One can convince oneself that
the symmetric part of the magnetic field $B_{(ij)}(S)$ can be written
as the functional derivative of this  functional $W[S]$
\begin{equation}\label{eq:symmag}
\frac{\delta}{\delta S_{ij}(x)}{W}[S] = \frac{1}{8\pi^2}B_{(ij)}(x)~,
\end{equation}
and thus the canonical transformation that eliminates the $\theta$-dependence
from the Hamiltonian can be represented
in the same from as (\ref{clctr}) with the nine gauge fields $A$ replaced by
the six unconstrained fields $S_{ik}(x)$.

\section{Concluding remarks}

We have explored the question of $\theta$-independence of
classical unconstrained $SU(2)$ gluodynamics in order to
build the basis for passing to the quantum level.
We have shown that the exact projection of $SU(2)$ gluodynamics to the
reduced phase space leads to an unconstrained system whose classical
equations of motion are consistent with the original degenerate theory
in the sense that they are $\theta$-independent.
The crucial point is that the fulfillment of this condition is due to
properly taking into account the Bianchi identity for the magnetic field.
As a consequence of the independence of the classical equations
of motion of the gauge invariant local fields, the parity odd term
in the Yang-Mills action is a total divergence of some gauge invariant
current, in contrast to the original unconstrained theory, where
it was the total divergence of the gauge variant Chern-Simons current
$K_\mu$. The explicit construction of the gauge invariant
current in the unconstrained theory remains a topic for further
investigation.
Furthermore, to deal practically with such a complicated nonlocal
Hamiltonian as (\ref{eq:uncYME}) one would have to use some approximation,
because the exact solution to equation (\ref{vecE}) is unknown.
Implementing the one or another approximating solution, it is desirable
to be consistent with the $\theta$-independence of classical theory.

\section*{Acknowledgements}

It is a pleasure to thank Z. Aouissat, A.N. Kvinikhidze, M.D. Mateev,
V.N. Pervushin, P. Schuck and A.N. Tavkhelidze
for illuminating discussions. D.M. and A.K. would like to thank
Prof. G. R\"opke for kind hospitality at the AG ''Theoretische
Vielteilchenphysik'' Rostock where part of this work has been done
and the Deutsche Forschungsgemeinschaft for providing a stipendium.
H.-P. P. acknowledges financial support by the  Bundesministerium
f\"ur Forschung und Technologie.


\end{document}